\input harvmac

\skip0=\baselineskip
\divide\skip0 by 2
\def\tmpsp{\the\skip0}

\def\skipthis#1{{}}

\def\sp{\partial\!\!\!/}
\def\sa{a\!\!\!/}
\def\sb{b\!\!\!/}
\def\ss{s\!\!\!/}

\Title{\vbox{\baselineskip12pt\hbox{HUTP-00/A014}
\hbox{\tt hep-th/0005059}}}
{\vbox{\centerline{From Noncommutative Bosonization to S-Duality}}}

\centerline{{ \bf Carlos Nu\~nez}, {\bf Kasper Olsen} and {\bf Ricardo Schiappa}} 

\bigskip\centerline{\it Department of Physics}
\centerline{\it Harvard University}
\centerline{\it Cambridge, MA 02138, U.S.A.}
\bigskip
\centerline{\tt nunez, kolsen, ricardo@lorentz.harvard.edu}

\vskip .5in 
\centerline{\bf Abstract}
\vskip .1in
\noindent
We extend standard path--integral techniques of bosonization and duality 
to the setting of noncommutative geometry. We start by constructing the
bosonization prescription for a free Dirac fermion living in the noncommutative
plane ${\bf R}_{\theta}^2$. We show that in this abelian situation the
fermion theory is dual to a noncommutative Wess--Zumino--Witten model. The
non--abelian situation is also constructed along very similar lines. 
We apply the techniques derived to the massive Thirring model on
noncommutative ${\bf R}_{\theta}^2$ and show that it is dualized to a 
noncommutative WZW model plus a noncommutative cosine potential (like in
the noncommutative Sine--Gordon model). The coupling constants in the fermionic
and bosonic models are related via strong--weak coupling duality. This is
thus an explicit construction of $S$--duality in a noncommutative field theory.

\smallskip
\Date{May 2000}
\lref\connes{A. Connes, M. Douglas and A. Schwarz, {\it Noncommutative Geometry
and Matrix Theory: Compactification on Tori}, JHEP {\bf 9802} (1998) 003, 
{\tt hep-th/9711162}.}
\lref\krogh{Y-K. E. Cheung and M. Krogh, {\it Noncommutative Geometry from 
$D0$--branes in a Background $B$--field}, Nucl. Phys. {\bf B528} (1998)
185, {\tt hep-th/9803031}.}
\lref\sw{N. Seiberg and E. Witten, {\it String Theory and Noncommutative 
Geometry}, JHEP {\bf 9909} (1999) 032, {\tt hep-th/9908142}.}
\lref\coolman{S. Coleman, {\it Quantum Sine--Gordon Equation as the Massive
Thirring Model}, Phys. Rev. {\bf D11} 2088 (1975).}
\lref\mandel{S. Mandelstam, {\it Soliton Operators for the Quantized 
Sine--Gordon Equation}, Phys. Rev. {\bf D11} 3026 (1975).}
\lref\nunez{J. C. Le Guillou, E. Moreno, C. Nunez and F. A. Schaposnik, 
{\it Non--Abelian Bosonization in Two and Three Dimensions},
Nucl. Phys. {\bf B484} (1997) 682, {\tt hep-th/9609202}.}
\lref\burgess{C. P. Burgess and F. Quevedo, {\it Nonabelian Bosonization as 
Duality}, Phys. Lett. {\bf B329} (1994) 457, {\tt hep-th/9403173}.}
\lref\bq{C. P. Burgess and F. Quevedo, {\it Bosonization as Duality}, 
Nucl. Phys. {\bf B421} (1994) 373, {\tt hep-th/9401105}.}
\lref\chu{C-S. Chu, {\it Induced Chern--Simons and WZW Action in 
Noncommutative Spacetime}, {\tt hep-th/0003007}.}
\lref\furuta{K. Furuta and T. Inami, {\it Ultraviolet Properties of 
Noncommutative Wess--Zumino--Witten Model}, {\tt hep-th/0004024}.}
\lref\cdp{M. Chaichian, A. Demichev and P. Presnajder, {\it Quantum 
Field Theory on Noncommutative Space-Times and the Persistence of 
Ultraviolet Divergences}, Nucl. Phys. {\bf B567} (2000) 360, {\tt hep-th/9812180}.}
\lref\witten{E. Witten, {\it Non--Abelian Bosonization in Two Dimensions}, 
Comm. Math. Phys. {\bf 92} 455 (1984).}
\lref\polyakov{A.M. Polyakov and P.B. Wiegmann, {\it Goldstone Fields in 
Two Dimensions with Multivalued Actions}, Phys. Lett. {\bf B141} 223 (1984).}
\lref\petersen{P. Di Vecchia, B. Durhuus and J.L. Petersen, {\it The
Wess--Zumino Action in Two Dimensions and Nonabelian Bosonization},
Phys. Lett. {\bf B144} 245 (1984).}
\lref\moreno{E. F. Moreno and F. A. Schaposnik, {\it The Wess--Zumino--Witten Term
in Noncommutative Two--Dimensional Fermion Models}, JHEP {\bf 0003} (2000)
032, {\tt hep-th/0002236}.}
\lref\redlich{A. N. Redlich and H.J. Schnitzer, {\it The Polyakov String in
$O(N)$ or $SU(N)$ Group Space}, Phys. Lett. {\bf B167} 315 (1986).}
\lref\minwalla{S. Minwalla, M. Van Raamsdonk and N.Seiberg, {\it 
Noncommutative Perturbative Dynamics}, {\tt hep-th/9912072}.}
\lref\mark{M. Van Raamsdonk and N. Seiberg, {\it Comments on Noncommutative 
Perturbative Dynamics}, JHEP {\bf 0003} (2000) 035, {\tt hep-th/0002186}.} 
\lref\alec{A. Matusis, L. Susskind and N. Toumbas, {\it The IR/UV
Connection in the Non--Commutative Gauge Theories}, {\tt hep-th/0002075}.}
\lref\naon{C. M. Naon, {\it Abelian and Non--Abelian Bosonization in the Path
Integral Framework}, Phys. Rev. {\bf D31} 2035 (1985).} 
\lref\schwarz{A. Schwarz, {\it Morita Equivalence and Duality},
Nucl. Phys. {\bf B534} (1998) 720, {\tt hep-th/9805034}.} 
\lref\boris{B. Pioline and A. Schwarz, {\it Morita Equivalence and
$T$--duality (or $B$ versus $\Theta$)}, JHEP {\bf 9908} (1999) 021, 
{\tt hep-th/9908019}.}
\lref\schiappa{L. Cornalba and R. Schiappa, {\it Matrix Theory Star Products
from the Born--Infeld Action}, {\tt hep-th/9907211}.}
\lref\lorenzo{L. Cornalba, {\it $D$--brane Physics and Noncommutative 
Yang--Mills Theory}, {\tt hep-th/9909 081}.}
\lref\ganor{O. J. Ganor, G. Rajesh and S. Sethi, {\it Duality and
Non--Commutative Gauge Theory}, {\tt hep-th/0005046}.} 
\lref\rajesh{R. Gopakumar, J. Maldacena, S. Minwalla and A. Strominger, 
{\it S--Duality and Noncommutative Gauge Theory}, {\tt hep-th/0005048}.}

\newsec{Introduction and Discussion}

\noindent
Quantum field theories on noncommutative spaces has been a subject of
renewed interest since the recent discovery of its connections to string
and $M$ theories, see {\it e.g.} \connes\krogh\schiappa\sw\lorenzo\ and
references therein. From a string theory point of view, it was realized 
in these works that one can translate the effects of a large background
magnetic field into a deformation of the $D$--brane world--volume. Still,
one can envisage studying such theories from a purely quantum field
theoretic point of view. For example, perturbative aspects of such 
noncommutative field theories have been studied and have revealed a 
surprising mixing of the IR and the UV \minwalla\mark\alec. These phenomena
are directly related to the string theoretic origins of these theories, but
one would also like to know to which extent properties of quantum field 
theories on commutative spaces also arise in quantum field theories on
noncommutative spaces. This may be of some interest given that it is not 
always simple to extract quantum results from string theory, while we 
are used to do so in field theory.

One important feature of many conventional quantum field theories 
is that of duality. As an example, it follows from bosonization
in 1+1 dimensions \coolman\mandel\ that the Sine--Gordon model of a
single scalar field,
\eqn\sineg{\int d^2x \; \{ {1\over 2}\partial_\mu\phi \partial^\mu\phi+{\alpha_0\over
\beta^2}(\cos\beta\phi - 1) \},}
is dual to the massive Thirring model of a fermion field,
\eqn\thirring{\int d^2x \; \{ \bar{\psi}(i\gamma^\mu\partial_{\mu}+m)\psi-
{\lambda \over 2} j_{\mu}j^{\mu} \}.}
One can relate bosonic composite operators to fermionic ones and
vice--versa using the standard bosonization machinery. Of particular
interest to us in here is that the Sine--Gordon/Thirring model duality is a 
strong/weak coupling duality since the coupling constants of the 
two theories are related according to:
\eqn\couplings{{4\pi \over \beta^2}=1+{\lambda\over \pi}.} 

The purpose of this paper is to study the analog of this duality on a 
noncommutative spacetime. In order to do this, we begin by considering
bosonization on the noncommutative plane and will see how the bosonization
rules get generalized to this situation. This is done in section 2, where
we study the abelian bosonization of a free fermion field in two noncommuting 
dimensions, employing path--integral techniques \naon\bq\burgess\nunez. We
shall learn that the free fermion action is bosonized to a noncommutative 
$U(1)$ WZW--action. That the WZW term in the action is nonvanishing for a $U(1)$ valued
field is simply due to the noncommutativity of spacetime. In fact, the
procedure follows much as for the conventional non--abelian bosonization
\witten, and the rules are very similar both in the non--abelian and
noncommutative cases. Because of this, the non--abelian noncommutative
bosonization will be a simple standard extension of the abelian noncommutative
bosonization. In particular, the non--abelian free fermion action bosonizes 
to a noncommutative $U(N)$ WZW model.

A question that immediately arises is the following. In the abelian case,
the free fermion has a quadratic kinetic action and therefore
noncommutative and commutative descriptions should match. On the other hand
the commutative abelian fermion is dualized to a free scalar field theory,
apparently very different from a noncommutative $U(1)$ WZW model. The same
phenomena happens in the non--abelian situation, where the commutative
abelian fermion is dualized to a commutative $U(N)$ WZW model, again not
the same as a noncommutative non--abelian WZW model. As we shall see, the
noncommutativity in the free fermion action makes its appearance when we
gauge the global symmetries in order to implement the path--integral
duality techniques of \bq\burgess.

The fact that this happens raises some interesting possibilities for
future research. Indeed this is apparently establishing some sort of
relation between commutative and noncommutative WZW models, and one could
interpret this as a different version of the Seiberg--Witten map between
commutative and noncommutative descriptions of the Born--Infeld action
\sw. It would be very interesting to find a string theory realization of
this field theoretic scenario, and try to understand this relation between
WZW models from a kind of $B$--field point of view.

After understanding the free fermion we proceed in section 3 to
interacting theories, with the goal of realizing $S$--duality for
noncommutative field theories. We shall see that the massive Thirring model 
on noncommutative space is dual to a WZW model plus a noncommutative cosine 
potential. The usual relation \couplings\ between the coupling constants of 
the dual theories continues to hold in the noncommutative case, thus
realizing an example of $S$--duality. Observe that in here the knowledge of 
the bosonization rules for the noncommutative free fermion plays a central
role, as they allow us to derive the noncommutative duality in very simple
steps. Indeed they allow for a full and explicit quantum construction of
$S$--duality in this noncommutative setting.

Understanding duality in noncommutative field theory, one could hope to
gain some starting grounds in order to try to match these results to a
string theory description. This is not a clear task, however. On one hand
the results in \sw\ are derived at the CFT disk level, so that one can not
assume that an $S$--duality in noncommutative field theory will translate
to a string theory $S$--duality. On the other hand, we are dealing with
simple bosonic theories which do not have immediate brane
realizations. After this paper was concluded, two pre--prints appeared that
describe $S$--duality in noncommutative gauge theories \ganor\rajesh, and 
therefore have a closer connection to the string theory description \sw. What
we would like to stress from our work is that not only it provides a
construction which does not rely on any string theory connection, but it
also allows for an explicit and exact treatment.

Some words on notation. To study noncommutative bosonization in 
${\bf R}_{\theta}^2$, the underlying ${\bf R}_{\theta}^2$ will be labeled
by noncommuting coordinates satisfying $[x^{\mu}, x^{\nu}]= i
\theta^{\mu\nu}$. Here $\theta^{\mu\nu}$ is real and antisymmetric and so
in two dimensions one has $\theta^{\mu\nu}=\theta\epsilon^{\mu\nu}$. The
algebra of functions on noncommutative ${\bf R}_{\theta}^d$ can be viewed
as an algebra of ordinary functions on the usual ${\bf R}^d$ with the
product deformed to the noncommutative, associative star product,
\eqn\starp{\big( \phi_1 \star \phi_2 \big) (x)=e^{{i\over 2} \theta^{\mu
\nu} \partial^y_\mu \partial^z_\nu} \phi_1(y) \phi_2(z)|_{y=z=x} .}
Thus, we shall study theories whose fields are functions on ordinary
${\bf R}^2$, with actions of the usual form $S=\int d^2 x {\cal
L}[\phi]$, except that the fields in ${\cal L}$ are multiplied using
the star product. Moreover, for any noncommutative theory the quadratic 
part of the action is the same as in the commutative theory, since if $f$ 
and $g$ are functions that vanish rapidly enough at infinity,
\eqn\quadac{\int d^d x  f \star g =\int d^dx fg .}

\newsec{Bosonization on Noncommutative Space}

\noindent
In this section we derive the bosonization rules of the free fermion action on a
two--dimensional noncommutative space \cdp, using the path--integral approach
described in \burgess\nunez. Our derivation is carried out for the abelian
case (and will largely follow \nunez) but as we shall see it generalizes
immediately to the non--abelian case \witten. In \witten\ it was
shown that a conventional free fermionic theory in 1+1 dimensions is
equivalent to a bosonic theory which is the WZW model. In our case the free 
fermionic theory is equivalent to a noncommutative version of the WZW
model, even in abelian case. 

\subsec{The Noncommutative WZW Model}

\noindent
Before looking at our specific problem, we start by taking the usual WZW model 
and define the noncommutative extension in the obvious way:
\eqn\ncwzw{S[g]={1\over 8\pi}\int_{\Sigma} d^2x\ (\partial_{\mu}g \star
\partial_{\mu}g^{-1})-{i\over12\pi}\int_B d^3x \epsilon^{ijk}\
(g^{-1}\star\partial_ig\star g^{-1}\star\partial_jg\star
g^{-1}\star\partial_kg).}
If one would like to extend this action to the non--abelian situation, one
simply needs to include a trace over the algebra. This theory has been 
discussed in a number of recent papers \moreno\chu\furuta. The manifold 
$\Sigma$ is parametrized by $(x^0,x^1)$ and is the boundary of the 
three--dimensional manifold $B$: $\partial B=\Sigma$. The $\star$--product 
on $B$ is the trivial extension of the product on $\Sigma$,
{\it i.e.} the extra dimension $x^2$ is taken to be commutative. The 
commutative non--abelian WZW model obeys the important Polyakov--Wiegmann 
identity \polyakov\petersen:
\eqn\pw{S[gh^{-1}]=S[g]+S[h^{-1}]-{1\over 4\pi}\int_{\Sigma}d^2x\
{\rm Tr}(g^{-1}\partial_+gh^{-1}\partial_-h) .}
The same identity holds with a $\star$--product in the noncommutative case
since the identity follows from using the cyclic property of the integral 
$\int A\star B =\int B\star A$ and from $g \star g^{-1}=1$. Here the group 
element $g$ of "noncommutative" $U(1)$ is
\eqn\group{g=e^{i\alpha}_\star=1+i\alpha-{1\over
2}\alpha\star\alpha
-{i\over 6}\alpha\star\alpha\star\alpha+\cdots .}
In the abelian commutative case, where $g=e^{i\alpha}$ without any $\star$--product,
the action \ncwzw\ of course reduces trivially to a free boson action $\int d^2x\
(\partial\alpha)^2$. However, the WZW action is nontrivial even for the
abelian noncommutative case, where $g=e^{i\alpha}_\star$.

\subsec{Definition of the Current}

\noindent
Using the path--integral approach to bosonization \nunez\burgess\ one
starts with the partition function of the free (abelian) fermion
theory:
\eqn\part{Z=\int{\cal D}\bar{\psi}{\cal D}
\psi e^{-\int\bar{\psi}\star i\sp\psi} .}
We want to show that this theory is equivalent to a given bosonic
theory. To prove that, one needs to show that the correlation functions
obtained from the two theories are equal. We therefore consider the
generating functional for these correlators as,
\eqn\cpart{Z[s]=\int{\cal D}\bar{\psi}{\cal D}
\psi e^{-\int\bar{\psi}\star (i\sp\psi+\ss)\star\psi} ,}
where $s_{\mu}$ is an external source. Due to the noncommutativity of our
problem, one might argue on where should one insert the source term, as for
instance $\bar{\psi}\star\ss\star\psi\neq\ss\star\bar{\psi}\star\psi$. As
we shall see later, \cpart\ is the definition we need in order to be able
to carry out the dualization procedure using the results for the fermionic
determinant in \moreno. So, we need to address the question of whether
\cpart\ is generating the correct conserved noncommutative current.

In the commutative fermi theory, the corresponding current is simply 
$j^{\mu}=\bar{\psi}\gamma^{\mu}\psi$ and this is conserved because of the 
equations of motion, which are $\sp\psi=0$ and
$\partial_{\mu}\bar{\psi}\gamma^{\mu}=0$. In the noncommutative case one 
finds similarly that the conserved current is:
\eqn\nccurrent{j^{\mu}=\bar{\psi}\star\gamma^{\mu}\psi=\bar{\psi}\gamma^{\mu}\psi+
{i\over 2}\theta^{\mu\nu}\partial_{\mu}\bar{\psi}\gamma^{\mu}
\partial_{\nu}\psi+\cdots .}
Now consider the following source--term in the partition function:
\eqn\crt{\int d^2x\ \bar{\psi}\star\ss\star\psi ,}
where $s_{\mu}$ is a source. It is not immediately obvious that this will generate
the correct correlation function, {\it i.e.} that
\eqn\source{{\delta\over \delta s_{\mu}(x)}Z[s]
=\langle j^{\mu}(x)\rangle,}
because of the infinite number of derivatives in the Moyal product. Let us
see that this is actually true. Because of \quadac\ one can remove for
example the last $\star$--product in \crt\ and then this integral equals,
\eqn\crtexp{\int d^2x\ \left(e^{{i\over 2} \theta^{\mu\nu}\partial^y_\mu
\partial^z_\nu}\ \bar{\psi}(y)\ \ss(z)|_{y=z=x}\right) \psi(x) .}
The first order term in $\theta$ can be written as,
\eqn\firstord{{i\over 2}\theta^{\mu\nu}\partial_{\mu}\bar{\psi}
\partial_{\nu}\ss\psi
={i\over 2}\theta^{\mu\nu}\partial_{\nu}[\partial_{\mu}\bar{\psi}\ss\psi]
-{i\over 2}\theta^{\mu\nu}\partial_{\mu}\bar{\psi}\ss\partial_{\nu}\psi .}
The first term on the RHS is a total derivative and so vanishes under the
integral sign. The second term on the RHS seems to have the wrong sign (and
the same appears to be the case with all higher--order odd terms in
$\theta$), since taking the functional derivative of the partition function 
with respect to $s_{\mu}$ will not lead to the current in
\nccurrent. However, the second term in \firstord\ also vanishes
identically because of antisymmetry of $\theta^{\mu\nu}$ and because of the 
following identity for Dirac fermions:
\eqn\diracprop{\bar{\chi}\gamma^{\mu}\psi=(\bar{\psi}\gamma^{\mu}\chi)^\dagger,}
which ensures that the current is real. Namely,
\eqn\vanish{
-{i\over 2}\theta^{\mu\nu}\partial_{\mu}\bar{\psi}\ss\partial_{\nu}\psi
=-{i\over2}\theta^{\mu\nu}(\partial_{\nu}\bar{\psi}\ss\partial_{\mu}\psi)^\dagger
={i\over2}\theta^{\mu\nu}(\partial_{\mu}\bar{\psi}\ss\partial_{\nu}\psi)^\dagger
={i\over 2}\theta^{\mu\nu}\partial_{\mu}\bar{\psi}\ss\partial_{\nu}\psi ,}
with the source $s_\mu$ being real. All higher--order terms with odd number
of $\theta$'s vanish for the same reason, {\it e.g.} the third order terms
is -- up to total derivatives -- of the form,
\eqn\thirdord{\theta^{\mu\nu}\theta^{\alpha\beta}\theta^{\gamma\delta}
\partial_{\mu}\partial_{\alpha}\partial_{\gamma}\bar{\psi}\ \ss\
\partial_{\delta}\partial_{\beta}\partial_{\nu}\psi}
and vanishes identically. This shows that the term in \crt\ does 
indeed generate the correct current.

\subsec{Path-Integral Derivation}

\noindent
In the following we will use the fact that the generating functional in 
equation \cpart\ is gauge invariant, {\it i.e.}
\eqn\ginv{Z[s]=Z[s^g] ,}
where under a gauge transformation the source transforms according to
\eqn\stransf{s_{\mu}\rightarrow s_{\mu}^g=g^{-1}\star s_{\mu}\star g
+g^{-1}\star\partial_{\mu}g .}
This invariance follows from the invariance of the measure under local
transformations of the fermion fields, {\it i.e.} transformations
$\psi\rightarrow e_\star^{i\alpha}\star\psi = g\star\psi$, together with
$\bar{\psi}\rightarrow \bar{\psi}\star e_\star^{-i\alpha}=\bar{\psi}\star g^{-1}$.
From \ginv\ we have:
\eqn\cpartb{Z[s]=\int{\cal D}\bar{\psi}{\cal D}\psi{\cal D}g\
e^{-\int\bar{\psi}\star (i\sp\psi+\ss^g)\star\psi} = 
\int{\cal D}g\ {\rm det}_\star(i\sp+\ss^g) ,}
where the last equality is obtained after integrating out fermions. Note
that the determinant is evaluated with respect to the $\star$--product (we
will shortly use the fact that this determinant was computed in \moreno). 
Introduce the connection
\eqn\bconn{b_{\mu}=s_{\mu}^g ,}
such that the field strengths of $b_\mu$ and $s_\mu$ are related according
to:
\eqn\fs{f_{\mu\nu}[b]=g^{-1}\star f_{\mu\nu}[s]\star g .}
We will choose a gauge where $b_+=s_+$, with $\Delta_{FP}$ being the
corresponding Faddeev--Popov determinant (we have 
$\Delta_{FP}={\rm det}_\star D_+[s_+]$, where
$D_+=\partial_++i[s_+,]_{\star}$).  This allows us to write \cpartb\ in the
form,
\eqn\cpartd{Z[s]=\int{\cal D}b_{\mu}{\rm det}_\star(i\sp+\sb)\
\delta[\epsilon_{\mu\nu}(f_{\mu\nu}[b]-f_{\mu\nu}[s])]\
\delta[b_+-s_+]\ \Delta_{FP} .} 
By introducing a Lagrange--multiplier 
field $\hat{a}$ that lives in the ``noncommutative'' $U(1)$ group with gauge transformation 
$\hat{a}\rightarrow \hat{a}^g=g^{-1}\star\hat{a}\star g$ one can write,
\eqn\cpartd{Z[s]=\int{\cal D}b_{\mu}{\cal D}\hat{a}\ 
{\rm det}_\star(i\sp+\sb)\ e^{\xi\int \hat{a}\star(f_{\mu\nu}[b]-f_{\mu\nu}[s])}\
\delta[b_+-s_+]\ \Delta_{FP} ,}
where $\xi$ is a constant which will be conveniently determined later. Now, 
make the following change of variables:
\eqn\change{\eqalign{
& s_{+} =i\tilde{s}^{-1}\star\partial_+\tilde{s} ,\cr
& s_{-} =is\star\partial_-s^{-1} ,\cr
& b_{+} =i(\tilde{b}\star\tilde{s})^{-1}\star\partial_-(\tilde{b}\star\tilde{s})
,\cr & b_{-} =(s\star b)*\partial_-(b^{-1}\star s^{-1}) .\cr}}
As we stated before, the fermion determinant for the noncommutative $U(1)$ theory has been
calculated in \moreno\ with the result that it is:
\eqn\stardet{{\rm det}_\star(i\sp+\sa)=\exp S_{WZW}[h\star g] ,}
where $a_+=h^{-1}\star\partial_+h$ and $a_-=g\star\partial_-g^{-1}$. The
action for the noncommutative WZW model on the RHS is given in equation
\ncwzw. With this result we can express the fermion determinant in terms of
the variables in \change:
\eqn\stardetb{{\rm det}_\star(i\sp+\sb)=\exp
S_{WZW}[\tilde{b}\star\tilde{s}\star s\star b] .}
The Jacobian for the change of variables $(b_+,b_-)\rightarrow
(b,\tilde{b})$ gives 
\eqn\bbjac{{\cal D}b_+{\cal D}b_-={\rm det}_\star D_+[\tilde{b}\star\tilde{s}]
{\rm det}_\star D_-[s\star b]{\cal D}b{\cal D}\tilde{b}
=\exp(\eta S_{WZW}[\tilde{b}\star\tilde{s}\star s\star b])
{\cal D}b{\cal D}\tilde{b} ,}
where we recall that the covariant derivatives $D_\pm$ are now in the
adjoint representation. Therefore the result for their determinant is the
same as for the fundamental representation but with an extra factor,
$\eta$, that accounts for the change in representation \redlich. This
factor can actually be computed to be related to the Casimir in the
commutative case, but as we shall never need it we simply leave it as
$\eta$. Furthermore, with this change of variables, one can write the 
$\delta$--function in \cpartd\ as:
\eqn\deltaf{\delta[b_+-s_+]={1\over {\rm det}_\star D_+[s_+]}\delta[b-1] .}
Combining these two results one obtains,
\eqn\parte{Z[s]=\int{\cal D}b{\cal D}\tilde{b}{\cal D}\hat{a}
\exp\left(S_{WZW}[\tilde{b}\star\tilde{s}\star s\star b]\right)
\exp\left[\xi\int d^2x\ \hat{a}\star(f_{+-}[b]-f_{+-}[s])\right]
\delta[\tilde{b}-1] .}
In the gauge $b_+=s_+$ we have $f_{+-}[b]-f_{+-}[s]=D_+[s_+]\star(b_--s_-)$ 
and so this gives,
\eqn\partf{\eqalign{
& Z[s]=\int{\cal D}b{\cal D}\hat{a}
\exp\left((1+\eta)S_{WZW}[\tilde{s}\star s\star b]\right)\cdot\cr
& \cdot\exp(-\xi\int d^2x\ D_+[s_+]\star \hat{a}
\star\left(is\star b\star\partial_-b^{-1}\star s^{-1})\right) .\cr} }
Note that the expression for the generating functional \cpart\ 
is gauge invariant, the transformation laws for $s,\tilde{s}$ being
$\tilde{s}\rightarrow \tilde{s}\star g$ and $s\rightarrow g^{-1}\star s$.
One further change of variables, from $\hat{a}$ to a group valued variable
$a$  is defined as follows: 
$D_+[\tilde{s}]\star\hat{a}=i\tilde{s}^{-1}\star(a^{-1}\star\partial_+a)\star\tilde{s}$
(note that $a$ is the bose field equivalent to the original fermi field and
will be invariant under gauge transformations). The Jacobian for the change of 
variables from $\hat{a}$ to $a$,
\eqn\jacobi{{{\rm det}_\star D_+[a\star\tilde{s}] \over {\rm det}_\star
D_+[\tilde{s}]},}
is, however, not gauge invariant. The trick \burgess\ is to use instead the 
following Jacobian obtained from the above by multiplying with (formally) one:
\eqn\invjacobi{{{\rm det}_\star D_+[a\star\tilde{s}]\over {\rm det}_\star
D_+[\tilde{s}]} {{\rm det}_\star D_-[s] \over {\rm det}_\star
D_-[s]} =\exp(\eta S_{WZW}[a\star\tilde{s}\star s])
\exp(-\eta S_{WZW}[\tilde{s}\star s]) .}
From this one finally obtains,
\eqn\cpartf{\eqalign{
& Z[s]=\int{\cal D}a{\cal D}b
\exp \Big( (1+\eta)S_{WZW}[\tilde{s}\star s\star b]
+\eta S_{WZW}[a\star\tilde{s}\star s]-\eta S_{WZW}[\tilde{s}\star s] \cr 
& +\xi\int d^2x\ 
\tilde{s}^{-1}\star a^{-1}\star\partial_+a\star\tilde{s}\star s
\star b\star\partial_-b^{-1}\star s^{-1} \Big) .\cr }}

We can now apply the Polyakov-Wiegmann identity \pw\ for the noncommutative
WZW model. Also we choose the up to now arbitrary value of $\xi$ to be 
$\xi=-{1\over 4\pi}(1+\eta)$. This gives the following result for the
partition function:
\eqn\cpartg{Z[s]=\int{\cal D}a{\cal D}b
\exp\left((1+\eta) S_{WZW}[a\star\tilde{s}\star s\star b]-
S_{WZW}[a\star\tilde{s}\star s]+S_{WZW}[\tilde{s}\star s]\right) .}
Now, make a change of variables $b\rightarrow
\hat{b}=a\star \tilde{s}\star s\star b$ with trivial Jacobian. Then the 
${\cal D}\hat{b}$--integration factors and it is just a normalization
contribution, so that one obtains the simpler expression,
\eqn\cparth{Z[s]=\int{\cal D}a\exp\left(-S_{WZW}[a\star\tilde{s}\star s]
+S_{WZW}[\tilde{s}\star s]\right) .}
One final change of variables with trivial Jacobian is 
$a\star\tilde{s}\star s\rightarrow \tilde{s}\star a\star s$; together with the
Polyakov-Wiegmann identity this leads to our final result for 
the bosonization of the free fermion action in equation \cpart\ (renaming $a$ as $g$):
\eqn\final{\eqalign{
&Z[s]=\int{\cal D}g\exp \Big[ -S_{WZW}[g]_\star \cr
& -{1\over 4\pi} \int d^2x\ (s_+\star s_--s_+\star g\star s_-\star g^{-1}
-ig^{-1}\star\partial_+g\star s_--is_+\star g\star\partial_-g^{-1}) \Big] .\cr }}
Our prescription for the noncommutative currents becomes:
\eqn\nccurrents{\eqalign{
& \bar{\psi}\star\gamma_+\psi\rightarrow {i\over 4\pi}g^{-1}\star\partial_+g ,\cr
& \bar{\psi}\star\gamma_-\psi\rightarrow 
{i\over 4\pi}g\star\partial_-g^{-1} .\cr}}

This derivation shows that -- in the abelian case -- the free fermion action on
${\bf R}^2_{\theta}$ is bosonized to the noncommutative $U(1)$ WZW
model. From this result it also follows what should be done for the non--abelian system. In the
non--abelian case, $g$ in equation \ncwzw\ belongs to the "noncommutative"
$U(N)$ group, {\it i.e.} $g=e^{i\alpha^aT^a}_\star$ and one should
therefore include the
ordinary trace over the $N\times N$ matrices in the appropriate places, as in
\pw. When this is done one immediately realizes that the above
derivation goes through without any changes, except of course that the
group elements now belong to $U(N)$ (note that the evaluation of the
noncommutative fermion determinant in \moreno\ also applies to the $U(N)$
case). This
shows that in the non--abelian case the free fermionic theory is dual to a 
noncommutative $U(N)$ WZW model.

With these results in hand, one is lead to make the following
observation. The free fermionic action (in both abelian and non--abelian
cases) in the noncommutative plane is the same as in the commutative plane,
due to its quadratic nature. On the other hand, the standard commutative
free fermionic action is equivalent to the commutative WZW model \witten,
and in the abelian case in particular it is equivalent to a theory of a free
scalar field. This shows an equivalence between commutative and
noncommutative WZW models, and the map between these two models might be
some version of the Seiberg-Witten map \sw\ between ordinary Yang--Mills 
theory and noncommutative Yang--Mills theory. It would be very interesting
to further explore and make more precise this relation.

\newsec{Noncommutative S--Duality}

\noindent
In this section we discuss a noncommutative version of the well--known duality
\coolman\ between the Sine--Gordon model and the massive Thirring
model. As we have just seen, the free fermion theory discussed in section 2
is dual to a noncommutative WZW model. The next step is to study an
interacting fermionic system, where the natural candidate is the Thirring
model. We shall see that we can dualize this theory in a straightforward
manner, given the results of the previous section. Moreover, we will
unravel a strong/weak coupling duality in the procedure.

\subsec{The Thirring Model}

\noindent
We consider the Thirring model with the usual quartic coupling:
\eqn\quadcoupl{S_\lambda=-{\lambda \over 2}\int d^2x(\bar{\psi}\star\gamma^{\mu}\psi)\star
(\bar{\psi}\star\gamma_{\mu}\psi) = -{\lambda \over 2}\int d^2x\ j^{\mu}\star j_{\mu} .}
In order to bosonize this term one can either use the bosonization
prescription directly (as we know how to bosonize the currents) or do it via a
"completing the square" type of prescription at the path--integral
level. The two methods obviously yield the same result and we simply use
the first. Using the bosonization recipe of section 2 one immediately obtains 
from equation \nccurrents\ that the four--fermion interaction \quadcoupl\ 
corresponds to the following term in the bose theory:
\eqn\quadbose{ -{1\over 2}\lambda\int d^2x\ 2(\bar{\psi}\star\gamma_+\psi)\star
(\bar{\psi}\star\gamma_-\psi)\rightarrow 
{\lambda\over 16\pi^2}\int d^2x\ g^{-1}\star\partial_+g
\star g\star \partial_-g^{-1} .}
This term is quartic and cannot be made quadratic. This is unlike the
commutative theory, where the bosonized current coupling term becomes an
extra contribution to the quadratic kinetic term in the scalar action. In
here, by including the quartic coupling \quadcoupl\ the resulting
noncommutative theory becomes significantly different from the
corresponding commutative theory by the introduction of an infinite series
of derivative terms in the $\star$--product. The bosonized Lagrangian 
therefore becomes:
\eqn\wessquad{S_{WZW}+
{\lambda\over 16\pi^2}\int d^2x\ g^{-1}\star\partial_+g
\star g\star \partial_-g^{-1}.}

To identify the bosonized fermion theory with a certain bose theory, one
still needs a canonically--normalized scalar variable \bq.  Recall that we 
are dealing with noncommutative $U(1)$ group elements, so that one needs to 
look at scalar fields $\Lambda(x)$ appearing as
$g(x)=e^{i\Lambda(x)}_\star$. From the WZW action \ncwzw\ one has the term,
\eqn\wzwquad{{1\over 4\pi}\int d^2x\
\partial_+\Lambda\partial_-\Lambda+\cdots ,}
while from the bosonized Thirring coupling \quadbose\ we get a term,
\eqn\thirrquad{{\lambda\over 16\pi^2}\int d^2x\ \partial_+
\Lambda\partial_-\Lambda+\cdots .}
By adding these two contributions we get the following kinetic term for the
scalar field:
\eqn\quadscalar{{1\over 4\pi}(1+{\lambda\over 4\pi})
\int d^2x\ \partial_+\Lambda\partial_-\Lambda ,}
and so the  canonically normalized scalar variable is:
\eqn\normscalar{\phi=\left[{1\over 4\pi}(1+{\lambda\over 4\pi})
\right]^{1/2}\Lambda .}
In particular, stability of the bosonic theory requires
$\lambda>-4\pi$. This result \normscalar\ will shortly turn out to be 
important in determining the relation between the couplings of the
noncommutative ``Sine--Gordon''\foot{Observe that we will not actually have a
simple noncommutative extension of the Sine--Gordon model due to the
extra term appearing in \wessquad.} and Thirring models.

\subsec{The Massive Thirring Model}

\noindent
We shall now turn to the fermion mass coupling. The relevant term is
\eqn\mass{S_m=m\int d^2x\ \bar{\psi}\star\psi .}
In order to extend the previous discussion to the massive Thirring model we 
follow the procedure outlined in \bq, as it applies to our
situation. Indeed one can bosonize the mass term by considerations of
chiral symmetry alone, and the discussion in \bq\ directly applies in here
as well (as one is considering global chiral symmetry for which the 
$\star$--product collapses to the standard product).

Free fermions are invariant under the global $U_A(1)$ axial
symmetry $\psi \to e^{i\alpha\gamma_5}\, \psi$ and $\bar{\psi} \to \bar{\psi}
\, e^{i\alpha\gamma_5}$, and one expects that the bosonic theory will share
such a symmetry \witten. On the other hand, under the duality procedure of the
previous section (see also \bq) one is gauging the global vector symmetry
of the free fermions and the axial symmetry will not survive quantization
due to the presence of the background gauge field. This axial anomaly in the
noncommutative plane was computed in \moreno,
\eqn\axialanomaly{\partial_\mu j^\mu_5 = -{1\over 2\pi}
\epsilon^{\mu\nu}\hat{F}_{\mu\nu} ,}
where the axial current is $j^\mu_5 = \bar{\psi} \star \gamma^\mu \gamma^5
\psi$ and $\hat{F}_{\mu\nu}$ is the noncommutative gauge field
strength. Because both initial and final theories (under duality) share the
same symmetry, it may seem odd that this symmetry is broken under the duality
procedure. The solution to make it manifest throughout the dualization
procedure is to include a transformation on the Lagrange multiplier as well
\bq. Indeed, recall that the Lagrange multiplier field $\Lambda$
appears in the initial gauged action as,
\eqn\lagrangemultipliergauged{\sim \exp \left( \int d^2x \; {1\over2\pi} \Lambda \star
\epsilon^{\mu\nu} \hat{F}_{\mu\nu} \right) .}
The observation is that if the field $\Lambda$ transforms as $\Lambda \to
\Lambda - \alpha$ under axial transformations, then the Lagrange multiplier term
\lagrangemultipliergauged\ will cancel the axial anomaly term \axialanomaly
as it appears in the path--integral, and chiral symmetry is made
manifest. In summary, what we have done is to nail down what is the correct 
transformation of the new field $\Lambda$ under chiral rotations, and this
is uniquely defined by the previous considerations (for details see \bq). 

The bottom line is that one can now proceed to deduce the bosonization of
fermionic mass terms from these transformation properties. Indeed, in
the presence of a mass term the fermi theory is no longer axial symmetric,
but the axial transformation rules nevertheless remain the same. As we
shall see in the following, these rules alone are enough information for a
unique determination of the bosonized composite operator that corresponds
to the fermionic quadratic term. For instance, we would like to bosonize a term
as $\psi_R^\dagger \star \psi_L$, which amounts to finding an appropriate
bosonic functional ${\cal F}(\Lambda)$ such that ${\cal F}(\Lambda) \equiv 
\psi_R^\dagger \star \psi_L$. Under global chiral transformations the
chiral mass term transforms as,
\eqn\chiralmass{\psi_R^\dagger \star \psi_L \to e^{-2i\alpha}
\psi_R^\dagger \star \psi_L ,}
and due to the $\Lambda$ chiral transformation rule just deduced, it
follows that the bosonic functional must satisfy,
\eqn\bosonizedmass{{\cal F}(\Lambda - \alpha) = e^{-2i\alpha}
{\cal F}(\Lambda),}
where we recall that due to the global character of the rotation, the
exponential involves no $\star$--product, even though the functional ${\cal
F}$ will be defined in terms of the field $\Lambda$ through a
$\star$--product. Indeed, it immediately follows that
\eqn\bosmass{{\cal F}(\Lambda) \propto e_\star^{ 2 i \Lambda}}
uniquely solves the functional equation (where the exponential is properly defined
with the $\star$--product). This whole procedure naturally follows from \bq\
as one is dealing with global chiral rotations. What all this amounts to, 
is that the mass term \mass\ bosonizes to:
\eqn\massbosonized{S_m = \int d^2x \; m\alpha_0 \cos_\star 2\Lambda ,}
where $\alpha_0$ is a constant associated to the zero point energy (as in
the commutative case \coolman), and the noncommutative cosine is defined
naturally by $\cos_\star \varphi = {1\over2} (e_\star^{i\varphi} +
e_\star^{-i\varphi})$. This story is very similar to the commutative
Sine--Gordon/Thirring duality. In particular, the coupling constant in the
bosonic theory is defined as $\beta$ and appears in the action exactly
through a cosine potential. In the noncommutative case, this would be,
\eqn\pot{\int d^2x \; \alpha \cos_\star \beta\phi}
where the field $\phi$ is the canonically normalized field. Finally, since
the canonically normalized field was 
$\phi=\sqrt{{1\over 4\pi}(1+{\lambda\over4\pi})}\Lambda$ we obtain the 
following $S$--duality property:
\eqn\sdual{{16\pi\over \beta^2}= 1+{\lambda\over4\pi} .}
This is exactly the same type of relation as that of the commutative theory 
\couplings. At first glance, one could worry that when we plug this
relation \sdual\ back in \wessquad\ there could be terms going like inverse
powers of $\beta$ which would spoil the strong/weak coupling
duality. However, this does not happen as one should also recall that the
scalar fields have to be canonically normalized according to
\normscalar. What therefore happens is that an expansion of \wessquad\ in
powers of the canonically normalized field $\phi$ will only produce
interaction terms proportional to positive powers of the bosonic coupling
constant $\beta$. From this we see that the strong/weak coupling duality between
the Sine--Gordon model and the massive Thirring model survives on a
noncommutative space, as we have just shown. 

It is known that the $T$--duality of string theory can be interpreted
as Morita equivalence in noncommutative geometry \schwarz\boris. One can
wonder if there could be a similar geometrical interpretation of string
theory $S$--duality in terms of noncommutative geometry. In here we have
given some first steps, by showing that one can construct quantum field
theoretical models in noncommutative space displaying $S$--duality.

\bigskip
\noindent
{\bf Acknowledgements:} We have benefitted from discussions/correspondence 
with C-S. Chu, L. Cornalba, K. Hori, A. Matusis, S. Minwalla, E. Moreno,
B. Pioline, F. Schaposnik and H. J. Schnitzer. CN is supported by
CONICET. KO is supported by the Danish Natural Science Research Council. RS
is supported by the Funda\c c\~ao para a Ci\^encia e Tecnologia, under the
grant Praxis XXI BPD-17225/98 (Portugal).

\vfill\eject
\listrefs
\end